*Note: This manuscript was originally submitted to Physical Review Physics Education Research on December 9, 2015, February 27, 2016, and July 22, 2017.*

# Measurement of the effect of member performances on team outcomes in two-stage exams

Hyewon Jang[1]

[1]*Office of Educational Innovation, Sejong University, Seoul, 05006, Republic of Korea*

## Abstract

Two-stage exams have been used in physics classrooms, yet no empirical study has examined the effect of individual member performances on team performance. Using direct measures of individual and team performances in two-stage physics exams, we study the effect of member performances on team outcomes. Our data reveal three results of interest. First, either the best or the mean of member performances significantly affects team performance, whereas the standard deviation of member performances has a weak effect. Second, member performances explain less than half the variance in team performances, indicating that team outcomes in two-stage exams are not determined simply by individual performances. Third, cross-validation shows that the best predictor of team performance is the best member's performance, highlighting the need to form balanced teams with high-scoring individuals distributed evenly throughout teams before team exams are conducted. As the first empirical analysis built on the input–process–output framework, this study provides researchers of two-stage exams with insight into how to investigate factors that can make team problem solving more productive in the team round.

## I. INTRODUCTION

From classrooms to business, we create teams and are required to work as team members. A team can be defined as a unit in which two or more individuals interact to achieve a common goal [1,2]. In the case of physics courses, scholars have successfully applied collaborative testing (i.e., two-stage exams) [3,4] after acknowledging that many students learn very little physics from traditional exams. The common approach is to add a collaborative component to the traditional individual exam so that students can discuss questions immediately after completing the exam. Our previous study shows that team scores in the team round represent not only knowledge transfer but also collaborative problem solving [5]. Arguably, member performance relevant to physics problem solving might affect the quality and quantity of team outcomes [6,7]. However, no study has illustrated how much of a team's outcome can be explained by member performance relating to physics problem solving.

Collaborative problem solving in two-stage exams has been shown to have a number of positive effects: improved performance [8], increased motivation to study [7,9], decreased test anxiety [10], a more positive relationship with classmates [9], increased retention [11], and increased positive perceptions of students and instructors [4,11]. However, in studies of two-stage exams, students have usually been randomly grouped or they have self-selected their teams [7,11]. Not surprisingly, students have reported concerns about team formation because they estimate that outcomes will be affected by students with high levels of knowledge or skills relevant to team tasks [12]. Indeed, little is known about the effect of member performance on team outcomes, whereas the positive effects of team exams on individual learning have been documented in research focusing on two-stage exams.

In this short paper, we investigate whether individual physics-problem-solving performance, as demonstrated in the individual round of two-stage exams, affects team outcomes. In this investigation, we examine how individual performance affects team performance, focusing on team members' knowledge and skills directly related to solving physics problems rather than on other factors, such as interpersonal interactions. Our research questions are as follows.

(1) Do team members' problem-solving performances in the individual round affect team outcomes in the team round?
- Does the highest member score affect the team score?
- Does the mean of team member scores affect the team score?
- Does the standard deviation of team member scores affect the team score?

(2) How much of team outcomes can be predicted by team members' problem-solving performance in the individual round?

## II. THEORETICAL FRAMEWORK

In this study, we refer to the input–process–output (IPO) framework developed by McGrath (1964) [13]. The IPO framework has been used in social psychology studies to analyze group behavior and performance for almost 40 years [2,14]. In particular, most research on team effectiveness has been substantially influenced by the IPO framework[15]. The framework considers input at three levels: the individual level (e.g., personal characteristics), the group level (e.g., group size), and the environment level (e.g., reward structure). There are two types of output: performance outcomes (e.g., the percentage of correct answers) and other outcomes (e.g., member satisfaction). The group interaction process mediates between input and output [2]. The input occurs before the interactive group process, and the output follows the group interaction. The framework assumes that “the input states affect group outputs via the interaction that takes place among members” (p. 317) [2].

Figure 1 presents the framework of this study, built on the IPO framework. In two-stage exams, individual knowledge and skills relevant to physics problem solving can be measured by the score obtained in the individual round. Team outcomes can be measured by the team scores. Group problem solving mediates between individual knowledge and skills, and team outcomes. In this study, we use three operational definitions of individual problem-solving performance within teams used in small-group research [6] as input variables: the best, mean, and standard deviation of team member scores. We assume that 1) the mean of individuals’ scores ($X_{mean}$) reflects the mean of member performances of the team, 2) the standard deviation of team members’ scores in a team ($X_{sd}$) represents the heterogeneity of individual knowledge and skills of a team relevant to solving test items, and 3) the best individual’s score in a team ($X_{best}$) represents the best member performance in a team.

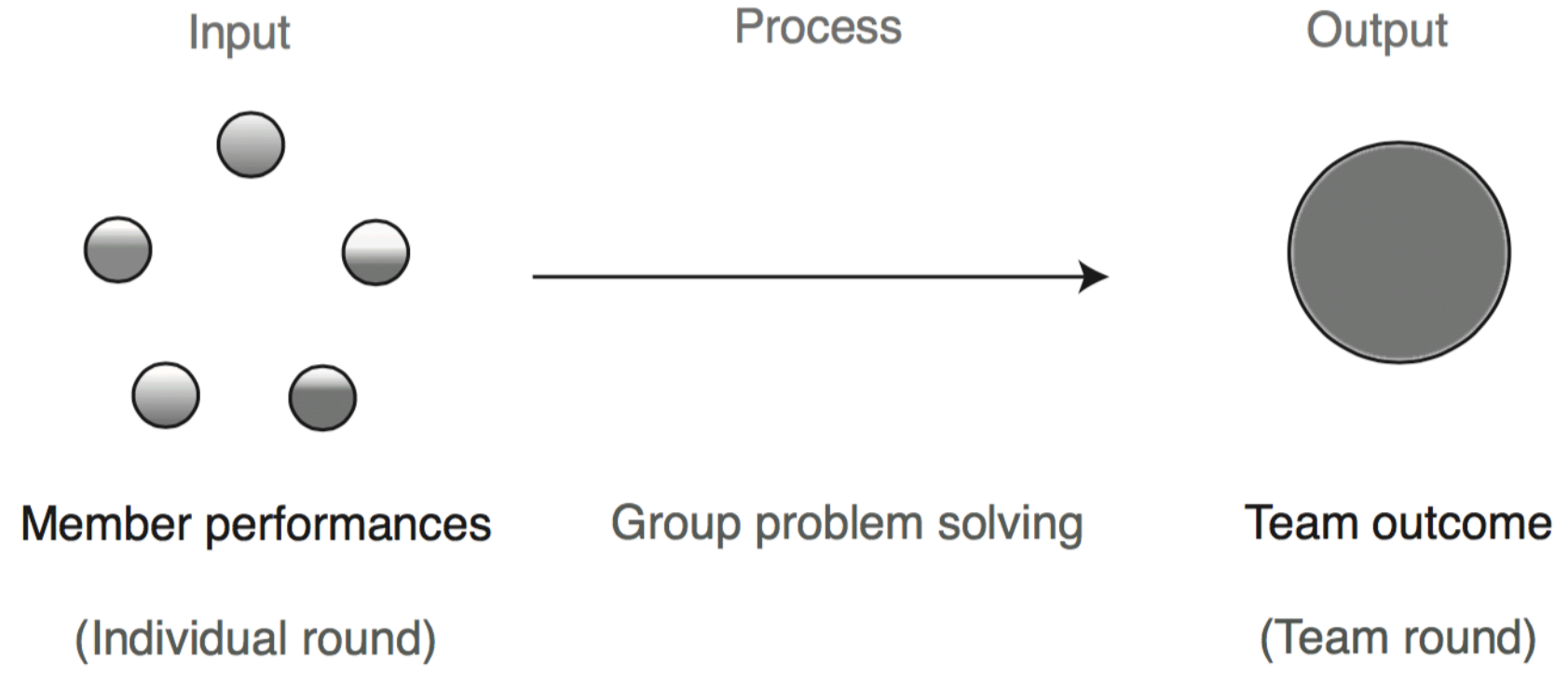

Fig. 1. Input–process–output framework for two-stage exams

# III. METHODS

## A. Participants

We collected two-stage exam data for 218 students ($N_{2014}$ = 67; $N_{2015}$ = 73; $N_{2016}$ = 78) in an introductory calculus-based mechanics class taught at Harvard University in 2014, 2015, and 2016. The three introductory courses were taught by the same instructors, mainly using two instructional methods: team-based learning [12] and peer instruction [16]. The classes met twice a week for a total of 6 hours of instruction per week. At the beginning of term, the gender, school year (freshman, sophomore, junior, or senior), Force Concept Inventory (FCI) scores, and class test results of all students were collected. A team consisted of four or five students with balanced gender, grades, academic experience (school year), background knowledge (as evaluated by the FCI pre-test score), and previous team makeup, following two rules: team members have complementary strengths (as measured by the FCI-pre-test score and grades) and women or minorities are not isolated [12].

## B. Two-stage exams

Two-stage exams were adapted from the Readiness Assurance Process for team-based learning [12]. The goals of the two-stage exams were to evaluate the students' understanding of physics concepts and problem-solving skills. In the two-stage exams, the students first solved test items individually and submitted their answers using an online response system [17]. After completing individual exams, they solved the same problems collaboratively with team members, and one team member submitted the answer on behalf of the team, using the response system[17]. Each assessment took approximately 90 minutes and had between eight and ten questions. Figure 2a illustrates four components of our approach—balanced teams, individual exams, team exams, and automated grading and instant feedback—compared with conventional two-stage exams consisting of individual and team exams in the literature [4,8]. In the individual round, each correct answer scored four points and no feedback was given. In the team round, a correct answer again received four points; however, if the initial response was incorrect, the team was given two more opportunities to submit a correct answer. All answers were automatically collected and graded by the system [17]. A correct answer given on the second try received two points, while a correct response given on the third attempt received one point (Figure 2b). The system revealed an answer only if a team answered correctly or failed to submit correct answers three times during team exams. For the two-stage exams conducted in 2014, 2015, and 2016, we gathered data comprising a total of 218 individual scores and 44 team scores.

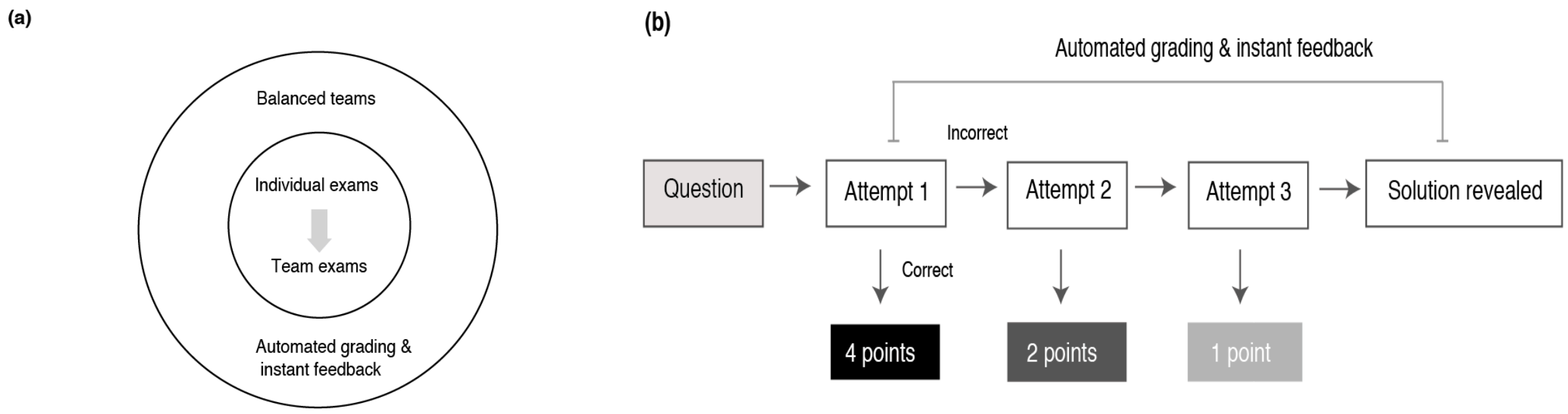

Fig. 2. (a) Core components of two-stage exams in the present study (large circle) vs. those of conventional two-stage exams (small circle); (b) schematic illustration of the team round in two-stage exams.

## C. Problem design

A total of 29 questions (seven open-ended and three multiple-choice questions in 2014; five open-ended and three multiple-choice questions in 2015; seven open-ended and four multiple-choice questions in 2016) were posed. The questions were either conceptual, computational, or estimation problems relevant to force, motion, and energy (see Appendix A). All questions were designed by a physics professor who had taught the subject for over 30 years and reviewed by physics educators in terms of the validity of content. To encourage collaborative problem solving, the questions were designed with a relatively high level of difficulty, so that even the strongest students would not score higher than 40% on the exam overall. This ensured that strong students would engage in collaborative problem solving.

## D. Analysis

First, we analyzed individual and team scores before and after the first collaboration in the team round. Second, we evaluated the correlation between individual performance and team outcome. Third, employing the IPO framework [13], we investigated the effect of member problem-solving performances on team outcomes using simple and multiple linear regression analyses. Before regression, we tested the correlation between variables that represent member performance in the individual round and selected uncorrelated variables as explanatory variables of a model.

In regression analysis, we assumed that the team scores ($Y$) are a sum of the weighted predictor variables and an error component ($\varepsilon$). The dependent variable ($Y$) contained team correctness after the first collaboration in the team round. Independent variables were the arithmetic mean ($X_{mean}$), standard deviation ($X_{sd}$), and best ($X_{best}$) of member performances. Here, we applied four regression models to measure the effects of the mean, best, and standard deviation of member performances on team outcomes.

Models 1–3 examine the effect of each of the best, average, and standard deviation of member performances in the individual round on team outcomes in the team round. We added Model 4 to include the effect of heterogeneity of member performances when comparing the effect of heterogeneity of member performances on team outcomes without controlling (Model 3) and while controlling (Model 4) for the mean of member performances. The regression models are expressed as

$$Model\ 1: Y = \beta'_0 + \beta'_1 X_{best} + \varepsilon,$$
$$Model\ 2: Y = \beta''_0 + \beta''_1 X_{mean} + \varepsilon,$$
$$Model\ 3: Y = \beta'''_0 + \beta'''_1 X_{sd} + \varepsilon,$$
$$Model\ 4: Y = \beta''''_0 + \beta''''_1 X_{mean} + \beta''''_2 X_{sd} + \varepsilon.$$

### E. Validation of models

The coefficient of determination ($R^2$) and residuals from fitted models can be analyzed and reported to validate regressions. However, residual evaluation does not reflect how well models can make predictions for data not already seen. We thus employed cross-validation (CV) techniques [18]. CV is a way of measuring the predictive performance of a statistical model and determining how the results of a statistical analysis will generalize to independent data. For CV, the data are divided into training and testing sets. Some of the data are used to train the model, and the remaining data are used to evaluate the performance of the model. Compared with residual evaluation, CV avoids over-fitting because the training sample is independent of the validation sample [19]. Estimates can be used to select the best model with information about the test error of the final chosen model. Here, we use 10-fold CV and leave-one-out CV to report the cross-validation residual sums of squares, which is a corrected measure of prediction error averaged across all folds. For details of the CV, see Kuhn and Johnson (2013) [18].

## IV. RESULTS

### A. Correlation between member performances in a team and team performance

Table I gives mean scores for individual and team exams. Having purposefully designed relatively difficult questions, the mean scores in the individual rounds range between 33% and 37%. Mean scores in the team round after the first collaboration were roughly twice the mean scores in the individual round, indicating that team scores doubled after the first collaboration. With two more trials, team correctness increased further to 82%–93%.

Table I. Mean scores in the individual round, $\langle S_i \rangle$, and team rounds after the first collaboration, $\langle S_{t1} \rangle$, and the third collaboration, $\langle S_{t3} \rangle$, for two-stage exams. The numbers in parentheses represent the standard deviation of reported values.

| Year | Participants | $\langle S_i \rangle$ | $\langle S_{t1} \rangle$ | $\langle S_{t3} \rangle$ |
|---|---|---|---|---|
| 2014 | 67 | 0.33 (0.06) | 0.57 (0.14) | 0.82 (0.10) |
| 2015 | 73 | 0.37 (0.09) | 0.63 (0.18) | 0.93 (0.06) |
| 2016 | 78 | 0.36 (0.16) | 0.68 (0.15) | 0.92 (0.07) |

To investigate the relationship between member performance and team outcomes, we plot team scores after the first collaboration for each test as a function of the mean, best, and standard deviation of individual scores in a team. Figure 3(a) shows that the average member performance in the individual round correlates with the team outcome after the first collaboration in the team round ($r = 0.60, p < 0.0001$). In Figure 3(b), the best of member performance in the individual round increases with the team outcome after the first collaboration in the team round ($r = 0.70, p < 0.0001$). Figure 3(c) shows that the standard deviation of member performance hardly correlates with the team outcome after the first collaboration ($r = 0.36, p = 0.016$). These results are consistent with the results of meta-analyses showing positive relationships between the medium and highest cognitive abilities in a team and the team performance [6].

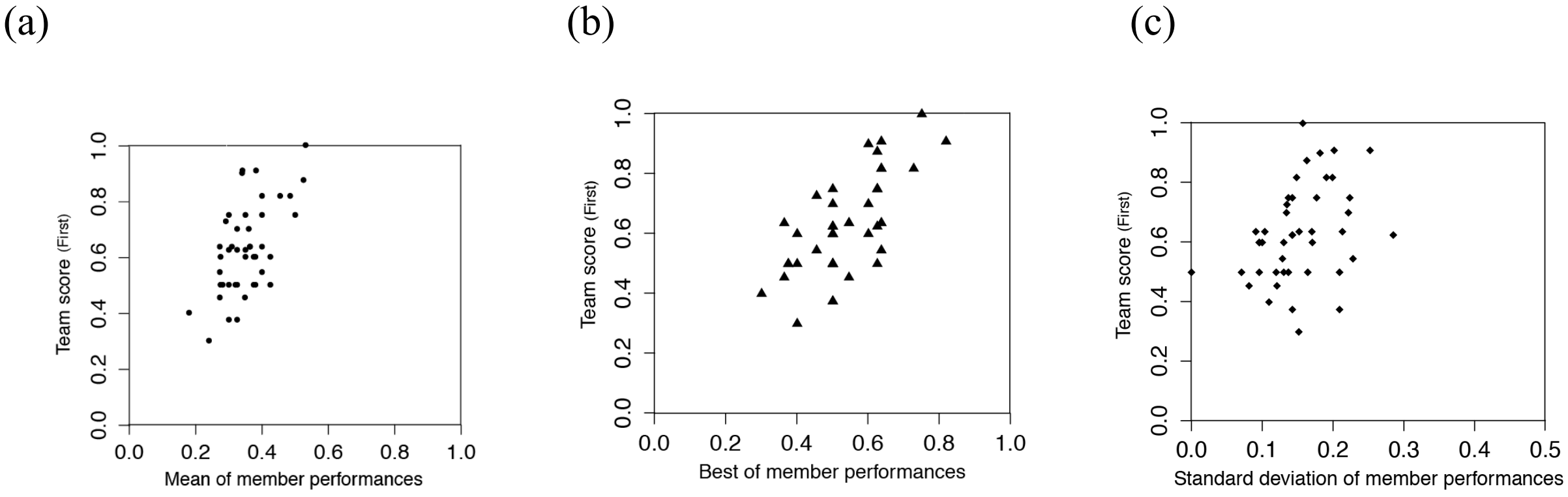

Fig. 3. Team scores after the first collaboration plotted for each test as a function of (a) the mean of member scores, (b) the best of member scores, and (c) the standard deviation of member scores of a team in the individual round.

## B. Effect of member performances on team outcomes

In Table II, regression results show that member performance in a team significantly affects team outcomes in the team round. With each additional point in the best individual performance of a team, we can expect the team performance after the first collaboration to increase by an average of 1 point according to Model 1. For each increase of 1 point in the mean individual performance of a team, the predicted team performance after the first collaboration increases by 1.29 points according to Model 2. Compared with the best or mean member performance, the heterogeneity measured by the standard deviation of member performances weakly affects team outcomes. In Table II, for a change of 1 point in the standard deviation of member performances, the average change in the mean of team performance is approximately 1.09 points according to Model 3. However, the effect of the standard deviation of member performances is less significant than either the effect of the mean or the effect of the best of member performances. When we control for the mean of member performances using Model 4, the effect is weaker. These findings show that either the mean or best individual performance significantly affects the team performance during team exams, whereas the standard deviation of member performances has a weak effect.

Table II. Standardized coefficients for linear regression models predicting team scores after the first collaboration in the team round. Model 1 controls for the best member score in a team. Model 2 controls for the average of member scores in a team. Model 3 controls for the standard deviation of member scores. Model 4 controls for both the average and standard deviation of member performances in a team. $^{a}$ $p < 0.1$; $^{b}$ $p < 0.05$; $^{c}$ $p < 0.0001$

| Model | 1 | 2 | 3 | 4 |
|---|---|---|---|---|
| Intercept | 0.1 | $0.17^{a}$ | $0.47^{c}$ | 0.09 |
| Best member score | $1.00^{c}$ | | | |
| Mean of member scores | | $1.29^{c}$ | | $1.18^{c}$ |
| Standard deviation | | | $1.09^{b}$ | $0.74^{a}$ |
| $R^2$ | 0.48 | 0.35 | 0.13 | 0.41 |
| RMSE | 0.117 | 0.1311 | 0.152 | 0.126 |
| P-value (F-statistics) | <0.0001 (39.36) | <0.0001 (22.92) | 0.01592 (6.312) | <0.0001 (14.31) |

## C. Variation of the team outcome explained by member performances

Here we report how much of a team outcome after the first collaboration can be explained by member performance. Table II shows that member performances explain less than 50% of the variance in the team outcome. Forty-eight percent of the variation in team performance after the first collaboration can be explained by the best of member performances, whereas 35% of the variation in team scores after the first collaboration can be explained by the average of member performances. Compared with the best or average of member performances, the heterogeneity measured by the standard deviation of member performances accounts for substantially less of the outcome; 15% of the variation in team correctness after the first collaboration can be explained by the standard deviation of team members' performances. However, the combination of the average individual scores in a team and the standard deviation of team members' scores seems to be a better predictor, accounting for 41% of the variation in team correctness after the first collaboration. In summary, approximately half of the variation in team scores is attributable to factors dependent on the individual scores of members constituting the team.

## D. Best model using the best score

Table III gives cross-validation (CV) estimates of prediction errors for each model based on 10-fold and leave-one-out CV. Results are CV residual sums of squares averaged across all folds for each model. A comparison of CV estimates of prediction errors of each model reveals that Model 1 has the smallest estimate, which means the prediction error for new data would be the smallest. The result suggests that the best of member performances can better predict team outcomes than the mean and the standard deviation of member performances in a team. In contrast, Model 3 has the largest estimate, which means the prediction error for new data would be the greatest. In other words, the standard deviation of member performances in the individual round is not an effective predictor of team outcomes. This finding reveals that the best score is the best predictor with which to predict team outcomes.

Table III. Cross-validation estimates of prediction errors for each model using 10-fold and leave-one-out cross-validation.

| Model | 1 | 2 | 3 | 4 |
|---|---|---|---|---|
| 10-fold | 0.015 | 0.019 | 0.024 | 0.018 |
| Leave-one-out | 0.014 | 0.018 | 0.024 | 0.017 |

## V. DISCUSSION

Researchers have reported positive relationships between the cognitive abilities of individual team members and team performance [6]; however, no study to date has directly tested the effect of member performance on team outcomes. This article's contribution is to provide the first empirical answer in the debate over the "effect of member performance on team outcome" in two-stage exams. Our findings show that either the mean or the best of member performances significantly affects team outcomes, with stronger members more likely to have higher team outcomes. Group problem-solving in the team round requires members to reach a consensus on the best solution. This consensus is built on the exchange of information [20]. Members with more physics knowledge would provide necessary physics knowledge, facilitate the dissemination of unshared information, and eliminate errors during group discussion [15,20-22]. This finding recommends that physics educators pay attention to the team make-up, especially when team scores are used to evaluate students' performance. Ideally, in two-stage exams, educators can form balanced teams using individual scores before the team round. To form balanced teams, it is recommended to ensure all groups have at least one higher scorer, because the best individual score can be used to predict the team outcome with the smallest prediction error.

We also found that the heterogeneity operationally measured by the standard deviation of member performances in the individual round weakly affected the team outcomes in two-stage exams, and much less strongly than the mean or best of member performances. There are notions that team member diversity is beneficial; i.e., a heterogeneous group outperforms homogeneous groups, especially in the case of intelligence-based tasks [6,23]. However, our finding shows that the heterogeneity of physics-problem-solving skills hardly predicts team outcomes, which aligns with the result that the standard deviation of member cognitive abilities was unrelated to team performance [6]. More studies are needed to confirm the effect of heterogeneity on team outcomes.

Our findings provide empirical evidence that unknown factors explain half of group performance. Indeed, 52% of the variance of team outcomes cannot be explained by member performances. Even though one team had the best student who scored 50% in the individual assessment, all team outcomes after the first collaboration were between 38% and 75%. When teams shared the same average individual score (e.g., 40%), they obtained different scores (e.g., ranging from 54% to 82%) in the team round. The IPO framework suggests that the group problem-solving process is one unknown factor. We ideally need to discover how group interaction mediates between individual member performance and team outcomes, including performance effectiveness. However, recording the group interaction in the classroom might affect students' group activities at the environment level [2]. Data in this study cannot be used to determine the relationship among the group interaction process, member performance, and team outcome and is limited in terms of the validity and reliability of problem sets. We encourage physics educators to test the effect of member performances on team outcomes using diverse problem sets.

Our research raises the question of whether the conventional approach used to measure team effectiveness is appropriate. Researchers of team effectiveness have measured team performance to represent team effectiveness [2,15,21]. However, the present study shows that high team performance is due not only to an effective team process but also to high member performance. The conventional approach of measuring team effectiveness by referring to team performance seems inappropriate for the study of what factors increase team performance for the same input. We might need a new approach that allows us to distinguish what factors promote group problem-solving for the same input. Future work should investigate latent factors accounting for the remaining 50% of variance in team outcomes for the same input. Such a study may explain why some teams have higher outcomes for the same input in the team round and explain differences between groups with higher and lower outcomes in two-stage exams.

## VI. CONCLUSION

Two-stage exams have been used in several disciplines by scholars who acknowledge that this type of assessment provides a powerful learning experience for students. However, empirical evidence showing the effect of member performance on team outcome in two-stage exams has been lacking to date. This article presented the first case of rigorous analysis. First, the best and mean of member performance affected team outcomes. Second, less than 50% of the variance of team outcomes can be explained by member performances. Third, the best member performance can better predict team outcomes than either the mean or standard deviation of member performances. It is recommended that in two-stage exams, physics educators form balanced teams by distributing the best individuals evenly throughout the teams.

## Acknowledgements

H.J. thanks the Mazur group at Harvard University, especially Prof. Eric Mazur for encouragement and support; Dr. Kelly Miller and Prof. Nathaniel Lasry for thoughtful and meaningful comments; Prof. Jung Bok Kim, Prof. Hyuk Joon Choi, and Prof. Hyewon Kim for support and discussions; Mazur group members for practical feedback and encouragement; and physics educators at the 2015 AAPT meeting. This research was partly supported by the Ministry of Education of the Republic of Korea and the National Research Foundation of Korea (NRF-2016S1A5B5A02025245).

Appendix A.

A1. Conceptual problem

Each of blocks in the figure below have the same inertia. Rank the configurations in increasing order of tension in the rope.

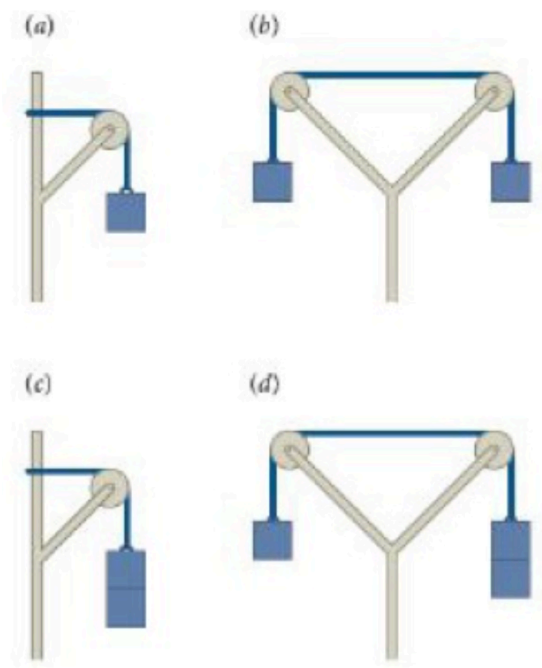

A2. Computational problem

After a hard day at the lab, you decide to go bungee jumping off a bridge. At the bottom of your jump, you can barely reach the ground. In that moment, you manage to hook another (identical) bungee cord to the ground. As you begin to rise back up to your starting position, you notice the second bungee cord is being stretched too. Because of this, you won't reach the starting height. What is the maximum height you can achieve, as a fraction of the starting height? Assume that the bungee cord is massless, and has an equilibrium length of zero (i.e. there is no slack in the bungee cord. It starts stretching as soon as you jump.)

A3. Estimation problem

Estimate energy dissipated when you clap your hands once[Joules]. Enter only the order of magnitude (For example, 2000J is entered as "3")